\documentclass[letterpaper,10pt]{article} 

\usepackage{opticameet3} 

\newcommand\authormark[1]{\textsuperscript{#1}}

\usepackage{amsmath,amssymb}
\usepackage[colorlinks=true,bookmarks=false,citecolor=blue,urlcolor=blue]{hyperref} 
\begin{document}

\title{Real-Time Diverse Fiber Sensing Multi-Event Detection using Phase OTDR Measurements}

\author{Konstantinos Alexoudis\authormark{1,2}, Jasper Müller\authormark{1}, Sai Kireet Patri\authormark{1}, Vincent A.J.M. Sleiffer\authormark{1}, Vishal Chandraprakash Rai\authormark{1}, André Sandmann\authormark{1}, Sander Jansen\authormark{1}, Thomas Bradley\authormark{2} and Chigo Okonkwo\authormark{2}}

\address{
    \authormark{1}Adtran Networks SE, Fraunhoferstraße 9a, 82152 Martinsried, Munich\\
    \authormark{2}High-Capacity Optical Transmission Laboratory, Eindhoven University of Technology, 5600 MB, Netherlands\\
}

\email{\authormark{*}konstantinos.alexoudis@adtran.com}

\begin{abstract}
We demonstrate an experimental phase optical time-domain reflectometry (OTDR) system capable of simultaneous detection and classification of various environmental events, such as wind-induced fiber movement, vehicle movement, and audio signatures, with real-time visualization.
\end{abstract}

\section{Overview}
The interest in distributed optical fiber sensing (DFOS) technology has surged in recent years, driven by a wide range of applications, including smart cities, environmental monitoring, and industrial automation. As infrastructure grows increasingly complex, real-time automated monitoring solutions, maintenance, remote security, and end-of-life monitoring have become crucial for optimizing operational expenditure (OPEX), efficiency, and ensuring resilience for critical infrastructure providers \cite{lu2019distributed}. DFOS systems based on phase (coherent) OTDR \cite{Lu2010jlt} can utilize the existing deployed fiber-optic infrastructure and benefit from inherent advantages of optical fiber, such as the long reach, low loss, and high bandwidth.

Event detection and classification are fundamental enablers for DFOS applications, as they allow for identifying and localizing critical events in real-time. DFOS systems rely on detecting changes in environmental parameters, such as temperature, strain, and vibrations, along the length of optical fibers \cite{lu2019distributed}. To provide actionable insights, the capability of classifying these events is essential. This enables a wide range of applications, including infrastructure monitoring, leak detection, and security surveillance such as detection of fiber tapping \cite{bradley2024}, where timely and accurate event detection can prevent failures and optimize operational efficiency \cite{lu2019distributed}.

Phase-OTDR systems produce large amounts of data that require real-time processing due to the continuous nature of the measurements. Several machine learning (ML)-based approaches to the analysis of these measurements have been proposed \cite{aktas2017deep, shiloh2018deep, wu2019one}. However, the training of these ML models requires large labeled datasets, which are very time- and cost-intensive to generate. Approaches to reduce the size of required datasets for ML model training have been proposed \cite{shi2022event}, and also event detection based on statistical methods have been proposed, reducing the reliance on data \cite{zhu2014vibration, jiang2018event}. 

We propose a demonstration of the detection and classification of multiple events in real-time using a phase-OTDR system. The system utilizes low-complexity event detection and feature extraction methods as well as ML algorithms. We present a user interface for real-time visualization of the measurements and event detection results. In this work, we aim to reduce the dependence on large labeled datasets by combining low-complexity event detection and identification \cite{mueller2024ACP} and lightweight ML-based classifiers to reduce the required data.

\section{Innovation}
This work showcases a unified approach for automated event detection and classification using an ML-based phase-OTDR system. The key innovations of this work include the simultaneous detection and localization of multiple events occurring at different positions along the fiber, the integration of ML algorithms for real-time event classification, and the demonstration of versatile application scenarios. Compared to existing methods, our approach enables the detection of multiple events concurrently in real-time. 

\begin{figure}[htb!]
    \centering
    \includegraphics[width=\linewidth]{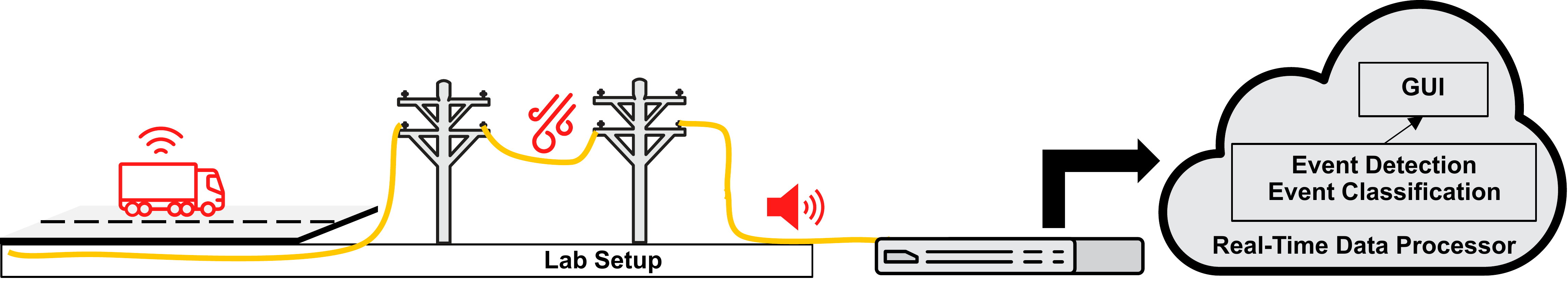}
    \caption{Small-scale experimental setup for event generation, including remote controlled car detection, fan-based wind generation, and audio speaker for acoustic event generation. A standard single-mode fiber, connected to the phase-OTDR, is deployed as a sensor in the setup, simulating buried and aerial fiber segments. The real-time data processor, consisting of containerized event detection, classification, and Graphical User Interface (GUI) applications, is deployed in the cloud.}
    \label{fig:setup}
\end{figure}

To validate our approach, we use a small-scale testbed modeling relevant scenarios for phase-OTDR systems (Fig.~\ref{fig:setup}). This setup allows for experimental verification of event detection and classification methods we propose, similar to practical demonstrations on deployed fibers \cite{azendorf2024}.

We utilize a combination of low-complexity event detection and identification methods \cite{mueller2024ACP} and ML models to analyze the OTDR measurement data. A real-time data processing pipeline enables the detection and localization of moving events as well as stationary events utilizing aggregated statistical metrics such as moving average and moving standard deviation. Detected events are classified using statistical features as input to a decision tree-based ML model. The model has been trained on a small labeled dataset of historical measurements from the testbed. Finally, the user interface visualizes the measurement data and detects events in quasi-real time.

\vspace{-1ex}
\vspace{-1ex}

\section{OFC Relevance}
At OFC, over recent years optical fiber sensors have been addressed in 2 main subcommittees; from exploiting fiber propagation effects, scattering and fiber devices in subcommittee D4: Fibers and Propagation effects, to the use of systems and digital signal processing in subcommittee S4: Optical Processing, Microwave Photonics and Fiber Sensing. Hence, DFOS and the increasing integration of ML into optical networks are active topics within the OFC community. Therefore, our demonstration presents a real-time, low-latency multi-event detection and classification system that operates with minimal training data. By addressing the challenges of large data requirements and computational complexity, our approach adds an advanced DFOS capability that improves practicality. We, therefore, believe this will be of interest to researchers, network operators, and equipment manufacturers seeking efficient, scalable, and integratable sensing solutions in optical networks.

\section{Objectives and Configuration of the Demo}

    \subsection{Experimental Setup}
    The experimental setup comprises a phase-OTDR device connected to a standard single-mode fiber and a cloud-based application with ML capabilities for event detection and classification. The miniaturized setup shown in Fig.~\ref{fig:setup} is designed to model real-world scenarios while providing the audience with interactive control over the environment. This is carried out as follows: the optical fiber is arranged into segments exposed to different types of events, starting with a segment of the fiber deployed overground and exposed to an acoustic signal from a loudspeaker to simulate different vibration events, such as mechanical digging. This is followed by a segment suspended between two masts and exposed to a fan, simulating environmental events such as winds. The last segment of the fiber is below a road with a remote-controlled robot car to detect vehicle movement. 
    
    \subsection{Multi-Event Detection Framework}
    A web application is deployed at the edge for data processing, event detection, and classification, enabling simultaneous localization of multiple events based on their unique characteristics (Fig.~\ref{fig:setup}). The server processes the signals in real time, relying on low-complexity event detection methods and a lightweight classification model trained on a small dataset.
    
    \subsection{Real-Time Monitoring of Environmental Events}
    Following the initial characterization, the system enables continuous monitoring of the infrastructure. Measurements can be configured and triggered via the control interface, enabling the detection and localization of new events. This real-time monitoring capability is crucial for infrastructure health monitoring, security surveillance, and environmental sensing applications, where timely and accurate event detection is essential.

\section{How the demonstration will be implemented}
Due to space constraints, the experimental part will be set up in Munich, Germany. For the audience, we will provide a live demonstration by remotely connecting to the setup. A live video will show the experimental phase-OTDR device connected to the optical fiber. The measurement data collected by the OTDR device will be sent to the cloud-based application and processed in real-time. A monitor at the OFC venue will display the GUI, showing the real-time data, detected events, and classification results. Attendees will be able to interact with the demonstration through the GUI, controlling simulation elements and observing the immediate effects on the system.

\section{How the demo will be presented to the attendees}
The live demonstration will feature real-time interactions and visualizations by interacting with the cloud-based fiber sensing application. The events monitored by the application will be generated by different sources along the fiber, such as a remote-controlled car, which will move along the road and generate vibrations along the fiber. The captured signal will be visualized on the display. Furthermore, events being detected by the event detection will be classified and displayed in the GUI, showcasing event detection and classification capabilities. The live demonstration will be presented to the attendees via screens at the venue. Attendees will interact with the cloud-based fiber sensing application through the GUI, allowing them to control various elements acting directly on the fiber, such as starting and stopping audio signals, operating the remote-controlled car, and toggling the fan in the remote experimental setup. As attendees manipulate these controls, events will be generated along the optical fiber in the laboratory setup. The resulting signals will be processed and visualized on the display in real time. The GUI will show the detected events and the event classification in real time, demonstrating the system's capabilities. Additionally, a live video feed of the experimental setup will be provided, allowing attendees to interact with the researchers in the laboratory, to observe the physical components and compare their interactions with the resulting measurements and event detection outcomes.

\section{How attendees might be able to interact with the demonstration}
Attendees will have control over all elements that can have an impact directly on the fiber. They can start different audio signals and see how the system detects different acoustic events at any time during the demo. They can also trigger the remote-controlled car to observe how different movement patterns affect the event detection and see how it's visualized on the display, and they can turn the fan on and off to see changes in wind-induced vibration. Furthermore, attendees can interact with the visualization software to explore the input data in more detail and engage with the system's functionalities.

\section{Acknowledgments}
This work has been partially funded by the German Federal Ministry of Education and Research in the project HYPERCORE \textbf{(\#16KIS2098)}. We also acknowledge the Bilateral Project \textbf{"DistraSignalSense"} between the Eindhoven University of Technology, The Netherlands, and Adtran Networks SE.

\end{document}